\documentclass[aps,prl,twocolumn,showpacs]{revtex4}
\usepackage{graphics,graphicx}

\usepackage{color}
\usepackage{wrapfig}
\usepackage{enumerate}
\usepackage{amssymb,amsmath}
\usepackage{bm}
\usepackage[normalem]{ulem} 

\begin{document}
\title{Magnetic and nematic orderings in spin-1 antiferromagnets with single-ion anisotropy}

\author{Keola~Wierschem$^1$, Yasuyuki~Kato$^2$, Yusuke Nishida$^2$, Cristian~ D.~Batista$^2$, Pinaki~Sengupta$^1$}
\affiliation{$^1$School of Physical and Mathematical Sciences, Nanyang Technological University, 21 Nanyang Link, Singapore 637371}
\affiliation{$^2$T-Division and CNLS, Los Alamos National Laboratory, Los Alamos, NM 87545}

\date{\today}

\begin{abstract}
We study a spin-one Heisenberg model with exchange interaction, $J$,  uniaxial single-ion exchange anisotropy, $D$, and Zeeman coupling to
a magnetic field, $B$, parallel to the symmetry axis. We compute the 
$(D/J,B/J)$ quantum phase diagram for square and simple cubic lattices by combining analytical and Quantum Monte Carlo approaches, and find a 
transition between  XY-antiferromagnetic  and ferronematic phases that spontaneously break the U(1) symmetry of the model.
In the language of bosonic gases, this is a transition between  a Bose-Einstein condensate (BEC) of single bosons and a BEC of pairs. Our work opens up new avenues
for measuring this transition in real magnets.
\end{abstract}
\maketitle


The unambiguous realization of BECs in laser cooled collections of cold atoms \cite{Anderson95,Davis95} triggered the search for more exotic states of matter and 
phase transitions that take place in bosonic gases. In parallel with these efforts, other experimental groups demonstrated that several aspects of the BEC quantum phase transition 
can also be measured in quantum magnets that are alternative realizations of bosonic  gases \cite{Nikuni00,Jaime04,Zapf06,Sebastian06a,Giamarchi08}. The main advantage of this alternative approach is that the much lighter boson mass 
leads to much higher transition temperatures, while the uniform character and well-defined temperature of quantum magnets are crucial for the study of quantum phase transitions. The main disadvantage is that the continuous symmetry associated with particle number conservation in atomic gases is always  an idealization for the case of quantum magnets
\cite{Sebastian06b}. However, many real magnets, for which the continuous symmetry breaking terms are much smaller than the ordering temperature, allow for measuring 
the universal behavior of  {\it continuous symmetry} breaking critical points over a large window of temperatures. 
The simplest example is the magnetic  field induced BEC quantum critical point (QCP) that is observed in several quantum magnets 
\cite{Nikuni00,Jaime04,Zapf06,Sebastian06a,Giamarchi08}. 

$S=1$ magnets can be mapped into gases of semi-hard core bosons via a  generalization of the Matsubara-Matsuda transformation \cite{Matsubara56,Batista04} that also maps the local magnetization into the boson density: $n_j = S^z_j +1$. In contrast to the  hard-core bosons associated with $S=1/2$ magnets,
 it is possible to study ``Hubbard-like''  bosonic gases with  on-site density-density  interactions because $n_j \leq 2$ \cite{Diehl10,Lee10,Bonnes11,Ng11,Chen11}.  
 Moreover, the semi hard-core constraint, $n_j \leq 2$, can be incorporated as an infinitely repulsive three-body on-site term that precludes
phase segregation when the two-body term is attractive. This situation is ideal for studying transitions between single-boson BECs and  BECs of pairs, 
whose counterparts in atomic physics are transitions between BECs of atoms and diatomic molecules (transitions between XY-magnetic and nematic orderings in the spin language). 

While the XY terms of the exchange interaction play the role of the kinetic energy, the Ising terms map into off-site density-density interactions.  On-site density-density interactions are generated by uniaxial single-ion anisotropy terms of the form $D (S^z_j)^2$. 
A magnetic field, $B$, parallel to the symmetry axis acts as a chemical potential because it couples to the total magnetization $M_z= \sum_j S^z_j$ that coincides with the total number of bosons up to an irrelevant constant.
Previous works have exploited this spin-boson mapping for studying spin supersolid phases of  $S=1$ Heisenberg models with uniaxial exchange and single-ion anisotropies \cite{Sengupta07a,Sengupta07b,Peters09}. The main purpose of this work is to study the 
quantum phase diagram (QPD) of the isotropic $S=1$ Heisenberg model when the single-ion anisotropy and Zeeman terms are included. This model is relevant for describing several Ni-based organic compounds as well as inorganic systems that are discussed at the end of this work.

Our $S=1$ model  is defined on a hypercubic lattice,
\begin{equation}
{\cal H}=J\sum_{\langle i,j\rangle} {\bm S}_{i} \cdot {\bm S}_{j} +D\sum_{i}\left(S_{i}^{z}\right)^{2}-B\sum_{i}S_{i}^{z},
\label{hamiltonian}
\end{equation}
and the antiferromagnetic (AFM) exchange coupling $J>0$ only connects nearest-neighbor sites $\langle i,j \rangle$.
Since an attractive on-site interaction is needed for pairing the bosons, we will only consider the 
$D<0$ case that corresponds to easy-axis anisotropy. 
%

\begin{figure}[t]
\includegraphics[width=7cm, trim= 30 30 30 30, clip]{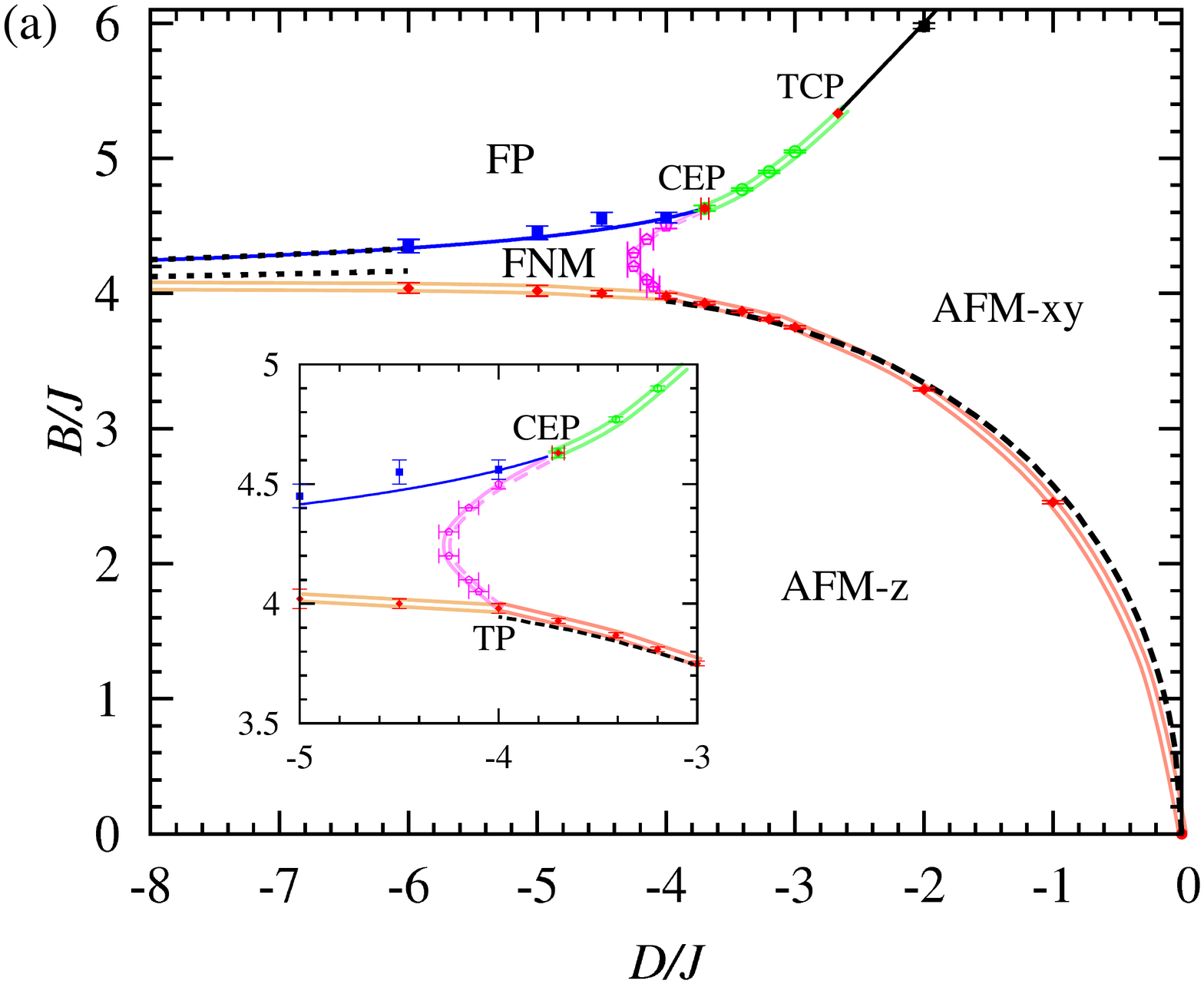}
\includegraphics[width=7cm, trim= 30 30 30 30, clip]{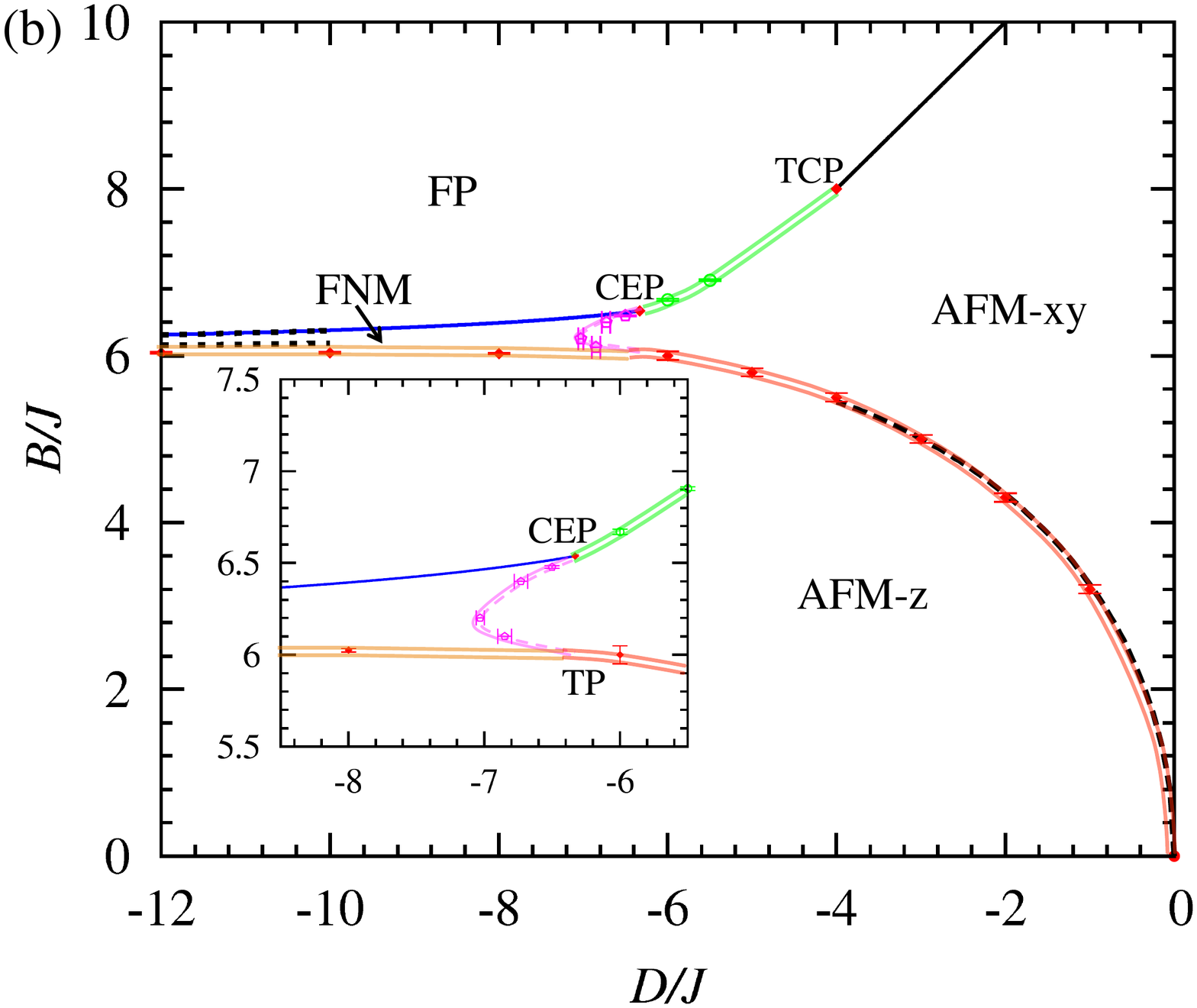}
\caption{
	Quantum phase diagrams in (a) square lattice and (b) cubic lattice.
	Double and single solid lines correspond to first and second order phase transitions respectively.
	Single solid lines are obtained by exact solutions while double solid lines are guides for the eye based on QMC  results.
	Insets: enlarged views of the FNM to AFM-xy transition.
	Dashed and dotted  lines are analytical solutions for the weak and strong anisotropy regimes respectively.
	}
\label{fig:fig1}
\end{figure}

%
Hamer {\it et al.}  computed the mean field QPD of ${\cal H}$ on a square lattice and obtained a single phase transition from AFM N\'{e}el (AFM-z) to the fully polarized (FP) state for large $|D|/J$ values~\cite{Hamer10}. However,  an intermediate nematic phase must exist in this regime according to the effective low-energy model, $\tilde {{\cal H}}$, that  is derived 
by expanding in the small parameter $J/|D|$ (strong coupling expansion in the bosonic language). The low-energy subspace, ${\cal S}$, is a direct product of the $S^z=\pm1$ doublets of each lattice site that are separated from the $S^z=0$ states by an energy gap $|D|$ \cite{Damle06}. 
$\tilde {{\cal H}}$ is obtained by applying a canonical transformation and projecting into  ${\cal S}$: $\tilde {{\cal H}}  = P_{\cal S} e^{\kappa} {\cal H} e^{-\kappa} P_{\cal S}$ ($\kappa$ is the anti-hermitian  generator of the canonical transformation and $P_{\cal S}$ is the projector on ${\cal S}$). If we use a pseudospin $s=1/2$ variable to describe each doublet, $S^z=2s^z$, we obtain the following expression for  ${\tilde {\cal H}}$ up to quadratic order in $J$:
\begin{equation}
\tilde {{\cal H}} =  
\sum_{\langle ij\rangle} {\cal J}^z  s^z_i  s^z_j +
{\cal J}^{xy}  (s^x_i  s^x_j+s^y_i  s^y_j)
- 2 B \sum_j s^z_j,
\end{equation}
with ${\cal J}^z = 4 J - J^2/D$ and ${\cal J}^{xy} =  J^2/D$.
${\tilde {\cal H}}$ is an $s=1/2$ XXZ model  whose QPD is well-known in any dimension. In particular, the mean field phase diagram is qualitatively and quantitatively correct
for spatial dimension $d\geq 2$. Since the case of interest corresponds to  effective strong easy-axis 
anisotropy, $|{\cal J}^z/{\cal J}^{xy}| \simeq 4|D|/J \gg 1$, the low field ground state has  N\'{e}el ordering that extends up to the spin-flop field, $B_{\rm sf} $, whose mean field value is 
$
B_{\rm sf} \simeq d\sqrt{({\cal J}^z + {\cal J}^{xy}) 
 ({\cal J}^z - {\cal J}^{xy})}/2.
$
The corresponding curve is the lower dotted line on the left of Figs.~\ref{fig:fig1}a and b. 
At $B=B_{\rm sf}$, the pseudospin variables flop from the N\'{e}el  state to a canted ferromagnetic state (${\cal J}^{xy} < 0$) whose  canting angle relative to the $z$-axis is given by $\cos{\alpha_{\rm sf}} = 2 \langle s^z_j \rangle=  \sqrt{({\cal J}^z+{\cal J}^{xy})/({\cal J}^z-{\cal J}^{xy})}$.  The effective  operator
for ${S^+_j}^2$ is ${\widetilde {{S^+_j}^2}}  = P_{\cal S} e^{\kappa} {{S^+_j}^2} e^{-\kappa} P_{\cal S} = 2 s^{+}_j $. This identity implies that the planar ferromagnetic ordering in the pseudospin variables or ``spin-flop" phase corresponds to ferronematic (FNM) ordering in the original $S=1$ spin variables, i.e., $\langle  s^{+}_j \rangle \neq 0$ implies  $\langle  {S^+_j}^2 \rangle \neq 0$. On the other hand, the effective operator for $S^+_j$ is equal to zero, ${\widetilde {{S^+_j}}}  = P_{\cal S} e^{\kappa} {{S^+_j}} e^{-\kappa} P_{\cal S} =0$, because of the following symmetry argument. Being odd under time reversal symmetry, ${\widetilde {{S^+_j}}} $ must be equal to a polynomial form that only contains odd terms in the  $s^{\nu}_l$ variables $\nu=\{x,y,z\}$. Such polynomial form must be odd under a $\pi$ spin rotation along the $z$-axis because $e^{i \pi \sum_l S^z_l} S^+_j e^{-i \pi \sum_l S^z_l} = - S^+_j $. Since 
$e^{i \pi \sum_l S^z_l} s^+_j e^{-i \pi \sum_l S^z_l} =  s^+_j $, the polynomial form must be equal to zero, implying absence of planar magnetic ordering in the large $D/J$ limit and confirming the FNM character of the intermediate phase.  The transition to the fully saturated state  is of second order in this regime and belongs to the BEC universality class in dimension $d+2$. A mean field treatment \cite{Hamer10} of the original Hamiltonian, ${\cal H}$, misses the second order fluctuations in $J$ (${\cal J}^{xy} =  J^2/D$) that stabilize the FNM phase.

The approximated value of the saturation field  is 
\begin{equation}
B_{\rm sat} (|D|/J \gg 1)  \simeq  \frac{d}{2}  ({\cal J}^z - {\cal J}^{xy})= d (2J-\frac{J^2}{D}),
\label{bsat}
\end{equation}
and the corresponding curve is the upper dotted line on the left of Figs.~\ref{fig:fig1}a and b. 
While Eq.~\eqref{bsat} is a good approximation for $B_{\rm sat}$ if $|D|/J \gg 1$, the exact curve $B_{\rm sat} (|D|/J \gg 1)$ can be computed analytically as long as the transition remains continuous. By solving the  two body problem of  diagonalizing ${\cal H}$ in the $M^z=N-2$ invariant subspace (two flipped  spins relative to the FP state)
we obtain the exact energy, $E_g(M^z=N-2)$, of the two bosons bound sate. The condensation of these pairs leads to the
FNM ordered state. If the transition is continuous, the exact value of $B_{\rm sat}$ is the field such that $E_g(M^z=N-2)=E_g(M^z=N)=N (J d+D-B)$. The resulting curve is shown as a full line in the $|D|/J \gg 1$ region of  Figs.~\ref{fig:fig1}a and b. 

The opposite limit, $|D|/J \ll 1$, is well described by a mean field treatment of the original Hamiltonian ${\cal H}$. The mean field N\'{e}el state 
$| \Psi_{\rm N} \rangle = \bigotimes_{j \in A} | 1\rangle_j  \bigotimes_{l \in B} | {\bar 1}\rangle_l$  is the most stable at low-enough fields. 
Here $A$ and $B$ denote the two sublattices, while  $| 1\rangle_j $, $| {\bar 1}\rangle_j $ and $| 0 \rangle_j$ are the eigenvectors of $S^z_j$ with eigenvalues $1$, $-1$ and $0$ respectively. In this regime there is a  spin-flop transition, but to a canted XY AFM (AFM-xy)  phase that is described by the mean field state 
\begin{equation}
| \Psi_{\rm sf} \rangle = \bigotimes_j  e^{i {\bm Q} \cdot {\bm r}_j}\sin{\theta} [\cos{\phi} | 1\rangle_j + \sin{\phi} | {\bar 1}\rangle_j ]  + \cos{\theta} | 0 \rangle_j,
\nonumber
\end{equation}
where ${\bm Q}$ is the AFM wave vector that has all the components equal to $\pi$. The optimal variational parameters, $\theta_0$ and $\phi_0$, are obtained by minimizing the mean field energy $\langle \Psi_{\rm sf}  | {\cal H} |\Psi_{\rm sf}  \rangle$. The dashed line on the right of Figs.~\ref{fig:fig1}a and b corresponds to  the spin flop curve $B_{\rm sf} (D/J)$ that results from  $\langle \Psi_{\rm N} | {\cal H} | \Psi_{\rm N} \rangle=\langle \Psi_{\rm sf} (\theta_0, \phi_0) | {\cal H} | \Psi_{\rm sf}  (\theta_0, \phi_0) \rangle$.

\begin{figure}[htp]
\includegraphics[clip,width=8.6cm]{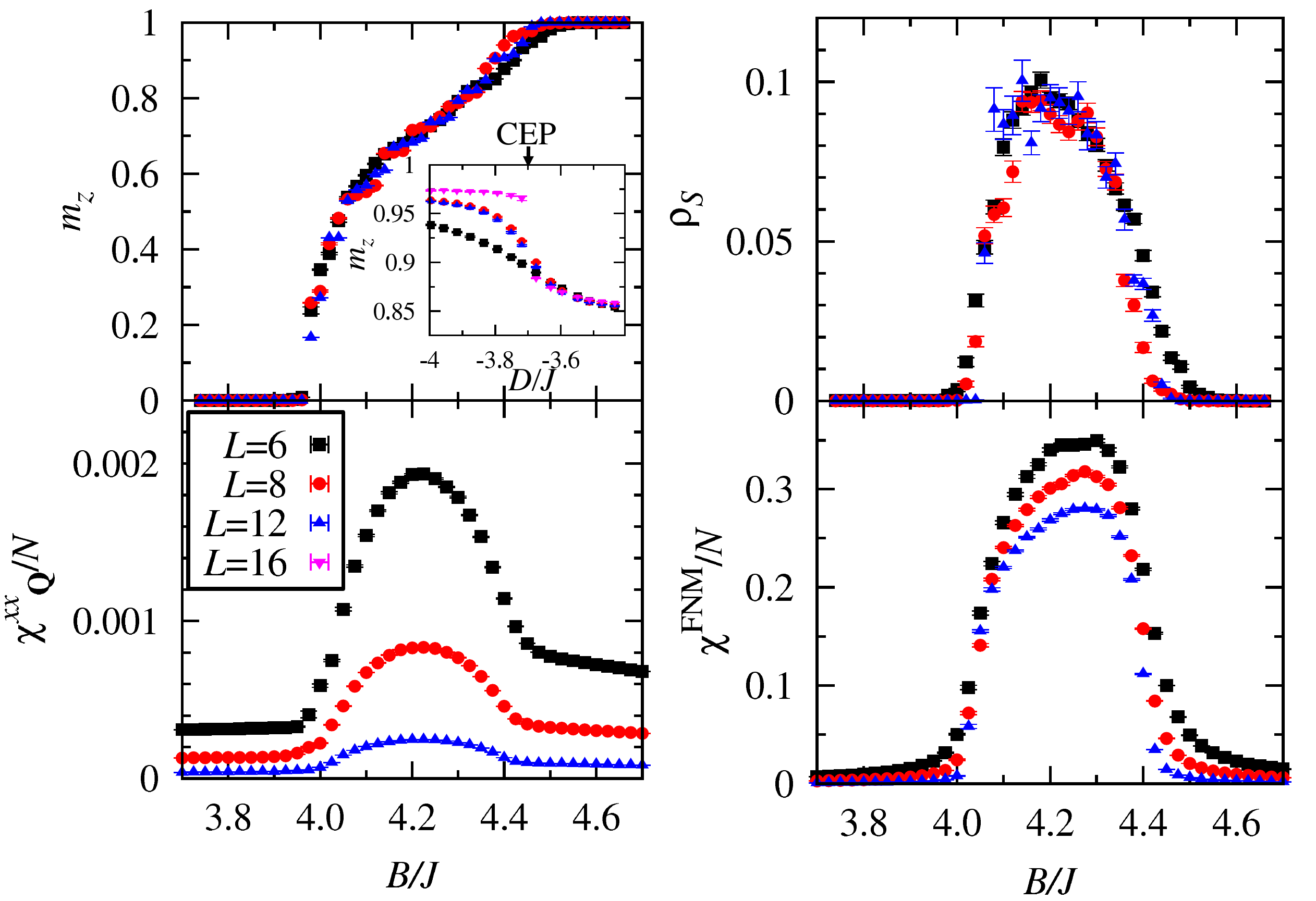}
\caption{
	Magnetic field sweep showing AFM-z to FNM to FP sequence of ground states for $d=2$ lattices of $L^2$ sites and periodic boundary conditions (PBC). Data shown are for $D/J=-5$ and inverse temperature of $\beta J=4L$. The inset shows $m_z$ along the line $B(D/J)$ where the  two-magnon ground state is degenerate with the  FP  state.}
\label{fig:fig2}
\end{figure}

\begin{figure}[htp]
\includegraphics[clip,width=8.7cm]{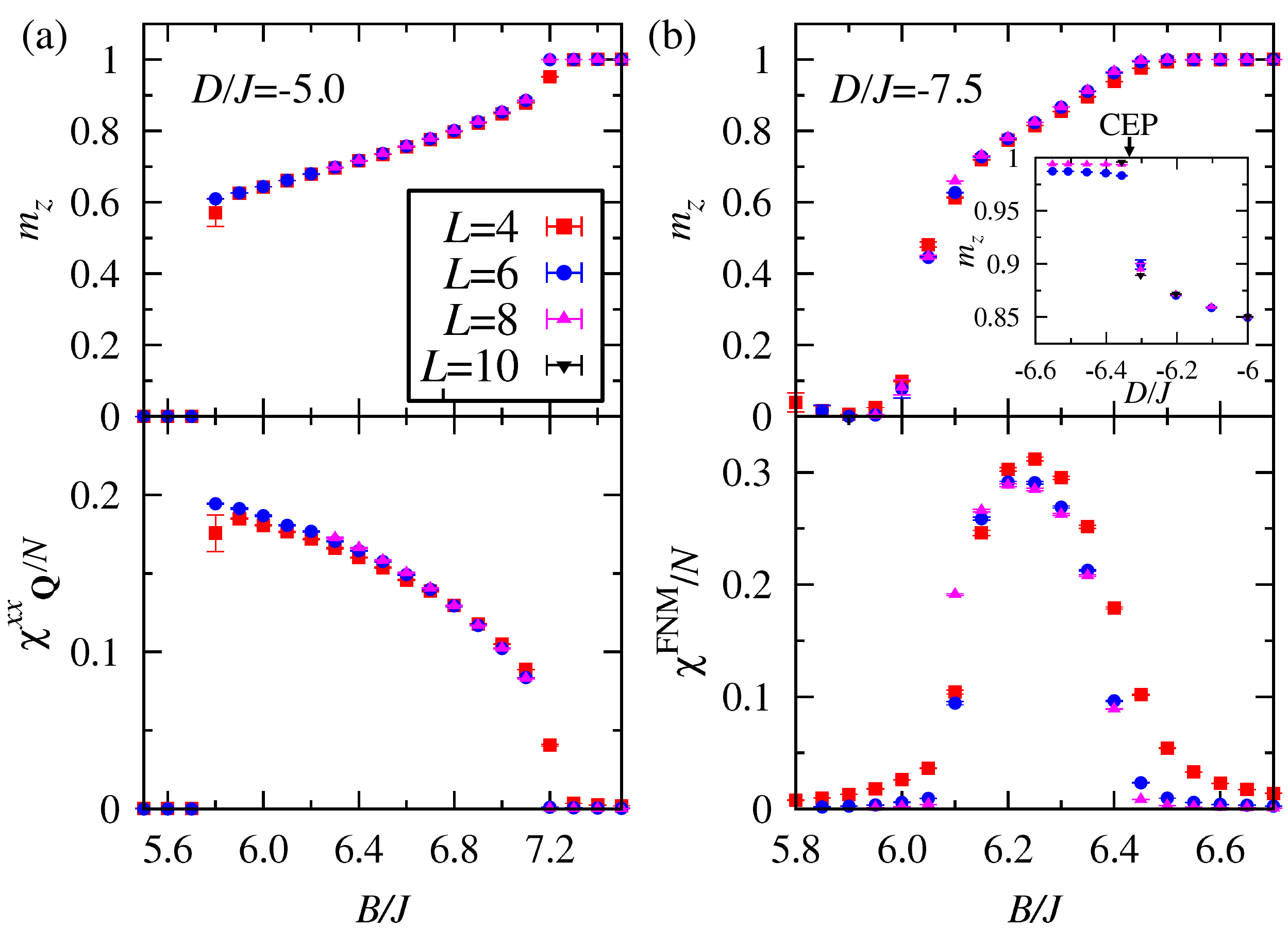}
\caption{ QMC results for $d=3$ lattices of $L^3$ spins  with PBC. The temperature is scaled with the systems size, $\beta J= 4L$, and the 
	Hamiltonian parameters are  (a) $D/J=-5$, (b) $D/J=-7.5$. Inset of (b) shows  $m_z$  along the line $B(D/J)$ where the  
	two-magnon ground state is degenerate with the  FP  state ($\beta J=6L$).}
\label{fig:fig3}
\end{figure}

For small $D/J$, the second order transition between the spin-flop and FP states belongs to the BEC universality class. 
The exact value of the saturation field in this regime is shown as
a  full line in the upper right region of Figs.~\ref{fig:fig1}a and b, and given by the equation
\begin{equation}
B_{0}  =  D +4 d J.
\label{bsat2}
\end{equation}
However, we know that the effective interaction between bosons should become attractive for $|D|>|D_{c1}|$. Therefore, the second order transition line described by Eq.\eqref{bsat2} should become of first order 
at a tricritical point (TCP) with coordinates $[D_{c1}/J, B_0(D_{c1}/J)/J]$ (see Figs.~\ref{fig:fig1}a and b).  
The region near the TCP is well described  by the Ginzburg-Landau (GL) free energy density,
\begin{equation}
f (\phi) = (B-B_{0}) |\phi|^2 + u |\phi|^4 + w |\phi|^6.
\label{GL}
\end{equation}
Here $\phi$ is the complex order parameter for the BEC of single bosons. 
$u$ and $w$ are the amplitudes of the effective two-body and three-body  interactions in the long wavelength (or continuum) limit. The field induced transition is continuous for repulsive $u>0$ and it happens at $B_{\rm sat}=B_0$ [see Eq.\eqref{bsat2}]. However, it is clear  from Eq.~\eqref{GL} that the transition becomes discontinuous for $u<0$. In this case, the transition field is $B_{\rm sat}=B_0 +  u^2/4w$ and the discontinuous change of
the boson density  is $\Delta m_z =\Delta |\phi|^2 = -  u/2w$ ($m_z=\sum_j \langle S^z_j \rangle/N$). The amplitude $u$ changes sign when the two-boson scattering length, $a_s$, diverges in $d=2$ ($a_s\to \infty$) and becomes equal  to zero in $d=3$ ($a_s =0$). These conditions determine the values of $D_{c1}/J$ in $d=2$ and 
$d=3$ respectively, that can be obtained by computing the effective interaction vertex in the long wave length and low frequency limits:
\begin{equation}
\Gamma_{\bm q}({\bm k},{\bm k'};\omega) = V_{\bm q}({\bm k}) + \int_{-\pi}^{\pi} \frac{d^dp}{(2\pi)^d} 
 \frac{V_{ {\bm q} - {\bm p}}({\bm k}) \Gamma_{\bm  p}({\bm k},{\bm k'};\omega)}{\omega - \epsilon_{{\bm k}+{\bm p}} - \epsilon_{{\bm k'}- {\bm p}} + i \delta}
\label{ladder}
\end{equation}
where  $V_{\bm q}({\bm k})= 2D+ \gamma_{\bm q}+(\sqrt{2}-2)(\gamma_{{\bm k} + {\bm q}}+\gamma_{\bm k})$,  $\epsilon_{\bm q} =  (2dJ+ \gamma_{\bm q})$ is the single boson dispersion at the TCP and $\gamma_{\bm q} = 2J \sum_{\nu} \cos{k_{\nu}}$. 
By solving Eq.~\eqref{ladder} for ${\bm q  =0}$, ${\bm k}={\bm k'}={\bm Q}$, and $\omega \to 0$,
we obtain $|D_{c1}|/J= 4 d/3$. 
A similar analysis cannot be applied to the point where the FP phase boundary  changes from second  to first order coming from the strongly anisotropic side $|D| \gg J$. Note that the FNM phase disappears right at this critical $D=D_{c2}/J$ point (see Fig.~\ref{fig:fig1}), while the magnetization vs field curve becomes discontinuous (see insets of
Figs.~\ref{fig:fig2} and~\ref{fig:fig3}). This discontinuity  indicates  that it is a critical end point (CEP) and the  effective GL theory is not applicable. Then, the coordinates of the CEP must be obtained from the quantum Monte Carlo (QMC) simulations that we describe below.

Our analysis of the two opposite regimes, $|D|/J \gg 1$ and $|D|/J \lesssim 1$, indicates that there is a transition between AFM-xy and FNM orderings in the intermediate region. We  use a QMC method with global updates \cite{Kawashima04} for studying this regime, because there is no small parameter for validating an analytical approach. Although ${\cal H}$ does not have a negative sign problem, 
standard QMC algorithms cannot  output the off-diagonal FNM correlator  because of a slowing down problem. 
Therefore, we use a novel multi-discontinuity algorithm \cite{Kato12b}, that is based on the directed-loop algorithm \cite{Syljuasen02} and 
eliminates the problem.

The different  phases  are  characterized by computing  the zero frequency  AFM-xy and FNM susceptibilities, 
\begin{eqnarray} 
	\chi^{xx}_{\bm Q} &=& {1\over \beta N}\sum_{i,j}
	\int_0^\beta 
	\langle  S^+_i(\tau)S^-_j(0) \rangle 
	e^{i{\bm Q}\cdot({\bm r}_i-{\bm r}_j)}
	d\tau,
	\nonumber \\
\chi^{\rm FNM} &=& {1\over \beta N}\sum_{i,j}\int_0^\beta \langle S^+_i(\tau)S^+_i(\tau)S^-_j(0)S^-_j(0) \rangle d\tau, \nonumber	
\end{eqnarray} 
where $N=L^d$ is the number of lattice sites. We also compute standard thermodynamic quantities, like the magnetization, $m_z$, and the 
spin stiffness, $\rho_s$ (response of the system to a twist in the boundary  conditions), that is obtained from
the fluctuations of  the world lines winding numbers along  the principal axes \cite{Pollock87}.

Figs.~\ref{fig:fig1}a and b include the $d=2$ and $d=3$ QPDs obtained from our QMC results. 
Figs.~\ref{fig:fig2} shows the four different observables computed as a function of $B/J$ for  $D/J=-5$ and different system sizes. 
Except for the FNM-AFM-xy transition, the phase boundaries of the first order phase transitions, 
shown in Figs.~\ref{fig:fig1}a and b, are  determined from the size dependence of the  discontinuity of the uniform magnetization and the corresponding  kink in the energy density.
These boundaries  agree very well with our analytical solutions in the limiting regimes  $|D|/J \gg 1$ and $|D|/J \lesssim 1$. 
Moreover, the QMC results indicate that the transition to the saturated state becomes of first order between the CEP and TCP that we discussed above. The coordinates of the TCP coincide with the exact values obtained by solving Eq.\eqref{ladder}.
The coordinates of the CEP obtained from our QMC simulations are $D_{c2}/J=-3.70\pm 0.03$ for $d=2$ and $D_{c2}/J=-6.33\pm 0.03$ for $d=3$.
Figs.~\ref{fig:fig3}a and b show that the magnetization curve has a small discontinuity for $D/J=-5$, while it is continuous for $D/J=-7.5$. 
The first order transition line for $|D_{c1}|<|D|<|D_{c2}|$ falls consistently above the curves given by Eqs.~\eqref{bsat} and  \eqref{bsat2}, as expected from
our GL analysis near the TCP.

\begin{figure}[t]
\vspace*{0cm}
\includegraphics[clip,trim=0cm 0cm 0cm 2cm,width=8.7cm]{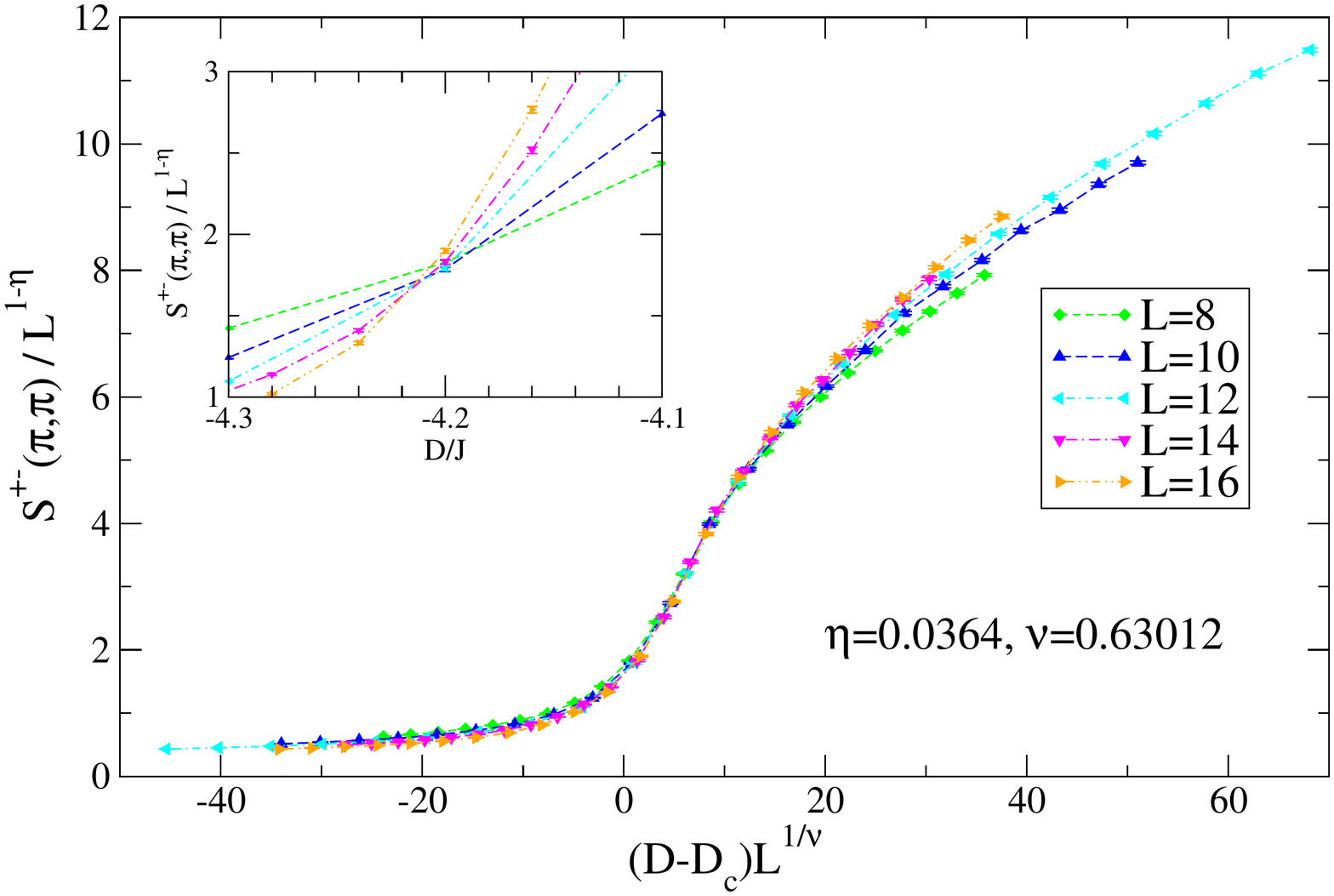}
\caption{
Finite size scaling of the QMC results for the AFM-xy structure factor, $S^{+-}({\bm Q})= \sum_{jl} S^+_j S^-_l e^{i {\bm Q}\cdot ({\bm r}_j -{\bm r}_l)}/N$, near the FNM to AFM-xy transition in $d=2$ with $B/J=4.1$ and $\beta=4L$. The values of the exponents for the Ising universality class in dimension $3=2+1$ are extracted from Ref.~\cite{Pelissetto02}. The common crossing point at $D_c/J=-4.22\pm0.02$ (see inset), together with the curve collapse in the critical region, 
suggests that the transition is continuous.
}
\label{fig:fig4}
\end{figure}

The transition from  FNM to AFM-xy ordering (double full-dashed line in Figs.~\ref{fig:fig1}a and b) spontaneously breaks the discrete symmetry of global spin rotation by $\pi$ along the $z$-axis. Consequently, if continuous, this transition should belong to the Ising universality class in dimension $d+1$.  The scaling analysis shown in Fig.~\ref{fig:fig4} indicates that this transition is most likely continuous away from the FP and AFM-z phases. However, our magnetization vs. field curves indicate that it becomes weakly first order near  the boundaries with these two phases, implying that the  upper end of the FNM to  AFM-xy phase boundary is a CEP,
while the lower end corresponds to a triple point (TP) at the junction of the FNM, AFM-xy, and AFM-z phases (see Fig.~\ref{fig:fig1}). 

The continuous or quasi-continuous nature of the FNM to AFM-xy quantum phase transition indicates that the single-boson condensate is continuously converted into a  condensate of
pairs (the condensate density is equal to the particle density $\rho= 1-m_z$ in the low-density limit). This observation implies that BECs of pairs and single-bosons coexist in a finite region of the AFM-xy phase that ends up at the phase boundary between the two phases where the single-boson BEC disappears completely: $\langle S^{+}_j \rangle=0$. Indeed, for $d=3$ and $D=-6.3J$, the size of the boson-pair, $\xi \simeq 0.77$, is three times shorter than the average inter-boson distance, $\rho^{-1/3}$, right below the saturation field ($\rho=1-m_z \simeq 0.1$).

The  AFM-xy and FNM orderings correspond to BECs of single bosons and pairs of bosons,
respectively. The shape of the phase boundary  opens the possibility of measuring magnetic field induced transitions between these two phases (see Figs.~\ref{fig:fig1}a and b).
Since a direct experimental detection of the spin-nematic order parameter can be rather challenging, our predictions for the quantum phase diagram and behavior of different thermodynamic properties are crucial for unveiling this  ordering in real magnets. While many $S=1$ magnets are described by ${\cal H}$ \cite{Becerra88,Wiedenmann89,Katsumata00,OConnor78}, it is vital to know what are the optimal ratios of $D/J$ for detecting the FNM ordering and characterizing the  FNM to AFM-xy quantum phase transition. Since most of these compounds are organic magnets, the $D/J$ ratio can be largely tuned as a function of pressure.Thus, knowing the appropriate range of $D/J$ values is necessary for detecting organic materials in which such a transition can be induced by pressure in  magnetic fields  that should be nearly  95\%  of the saturation field. 

Finally, we mention that field-induced spin supersolid states (coexistence of AFM-z  and FNM orderings) exist at least in the strongly anisotropic limit of ${\cal H}$ for  triangular \cite{Boninsegni05,Wessel05,Heidarian05} and face-centered-cubic  lattices \cite{Suzuki07}. Other exotic states have been reported for the kagome lattice \cite{Damle06}.
Ferronematic order has also been obtained for $S=1$ Heisenberg models that include biquadratic interactions \cite{Batista04,Harada02,Toth12,Lauchli06,Manmana11}.


\begin{acknowledgments}
We thank A. Paduan-Filho and J. Manson for valuable comments. This research used resources of the National Energy Research Scientific Computing Center (NERSC), which is supported by the Office of Science of the U.S. Department of Energy under Contract No. DE-AC02-05CH11231. Work at LANL was performed under the auspices of a J. Robert Oppenheimer Fellowship and the U.S.\ DOE contract No.~DE-AC52-06NA25396 through the LDRD program.
\end{acknowledgments}

\bibliography{ref}

\end{document}